\documentclass[12pt, a4paper]{article}

\usepackage[left=2.5cm, right=2.5cm, top=2.0cm, bottom=2.0cm]{geometry}
\usepackage{amsmath, amssymb, graphicx}
\usepackage[utf8]{inputenc}
\usepackage[T2A]{fontenc}
\usepackage[english]{babel}
\usepackage{amsthm}
\usepackage{color}
\usepackage{here}
\usepackage{hyperref}
\hypersetup{colorlinks,citecolor=blue,linkcolor=blue}

\newcommand {\tr} {\text{Tr}}
\newcommand{\fS}{\mathfrak{S}}

\title{Evolution of the holographic entanglement entropy quantities for composite quantum systems}
\author{Irina Aref'eva, Oleg Inozemcev, Igor Volovich}

\begin{document}

\maketitle

Steklov Mathematical Institute of Russian Academy of Sciences

119991, Russia, Moscow, Gubkina str., 8

\vspace{10mm} E-Mail: arefeva@mi-ras.ru, inozemcev@mi-ras.ru, volovich@mi-ras.ru

\vspace{4cm}
\begin{center}
	\textbf{Abstract}
\end{center}

\begin{center}
	\parbox{12cm}{
	Holographic entanglement entropy quantities for a three-party system are considered, namely tripartite information, total correlation and secrecy monotone. A holographic approach with Vaidya-AdS$_3$ is used to calculate time evolution of the entanglement entropy during non-equilibrium heating. We study time dependence of these three entropy quantities.	
 }
\end{center}

\newpage
\section{Introduction}

In recent years, various aspects of correlations and dynamics of quantum entangled states were considered in numerous papers, see for example \cite{OhyaVol}--\cite{KozMirTerVol} and refs therein. For a holographic approach to entanglement entropy and its time evolution, see for example \cite{RyuTak}--\cite{ArefInozVol:Hologr} and refs therein.

In \cite{ArefInozVol:Hologr}, we considered a holographic bipartite quantum system, one part of which consisted of a segment and the other of two segments. For such three-segment system, we investigated time dependence of the holographic mutual information between the two parts of the system. In that case, the three segments were not, obviously, included in the system in an equal manner. We found five fundamentally different types of time dependence of the holographic mutual information for that quantum system.

In this paper, we consider entanglement entropy quantities which are generalizations of the mutual information to a three-party system in which all segments are included in the same way. These quantities are the tripartite information, the total correlation and the secrecy monotone \cite{Horodecki}-\cite{Shir:Measures}. If entanglement entropy in these quantities is calculated holographically then one get holographic entropy quantities.

It is interesting that for the holographic total correlation and secrecy monotone we have found exactly the same five types of time dependence as for the holographic mutual information in \cite{ArefInozVol:Hologr}. For the holographic tripartite information, we have succeeded to find only four of these five types of time dependence.

\vspace{1cm}
\section{Holographic entanglement entropy in Vaidya-AdS$_3$}
We use a holographic approach to study evolution of an open system after a quantum quench accompanied by non-equilibrium heating process in the three-dimensional case. As the dual model describing time evolution of the entanglement during such a process, we consider a Vaidya shell collapsing on a black hole \cite{Balasub:Hol_Therm,ArefVol:Photo}. The collapse of this shell leads to the formation of a heavier black hole, which corresponds to a temperature increase. The initial thermal state is defined as the horizon position $z_H$, and the final state, as the horizon position $z_h$, $0<z_h<z_{H}$. 

The corresponding Vaidya metric defining the dual gravitational background consists of two parts and is given by
\begin{align}
	\label{Vm1a}
	v<0: ~~~ ds^2 = \frac{1}{z^2}\left[ -\left(1-\frac{z^2}{z_H^2}\right)dv^2 - 2\,dv\,dz + dx^2 \right], \\
	\label{Vm2a}
	v>0: ~~~ ds^2 = \frac{1}{z^2}\left[ -\left(1-\frac{z^2}{z_h^2}\right)dv^2 - 2\,dv\,dz + dx^2 \right]. 
\end{align}
The two parts are glued along the position of the shell $v=0$.

The initial and final temperatures are
\begin{equation}
	T_i=\frac{1}{2\pi z_H},\hspace{1cm} T_f=\frac{1}{2\pi z_h}.
\end{equation}	

To define the holographic entanglement entropy \cite{RyuTak,HubRangTak} corresponding to the simplest system, i.e. one segment, we have to find a geodesic in the metric \eqref{Vm1a}, \eqref{Vm2a} anchored on this segment at a given instant. The action for the geodesic connecting two points $(t,-\ell/2)$ and $(t,+\ell/2)$ on the boundary is given by
\begin{equation} \label{action1}
	S(\ell,t)=\int\limits_{-\ell/2}^{+\ell/2}\frac{dx}{z}\,\sqrt{1-2v'z'-f(z,v)v^{\prime 2}},
\end{equation}
where $t\geq0$, $\ell>0$, $z=z(x)$, $v=v(x)$ and
\begin{equation} \label{BC}
	z(\pm \ell/2)=0,\,\,\,\,v(\pm \ell/2)= t.
\end{equation}
The symmetry of the problem implies that
\begin{equation}
	z'(0)=v'(0)=0.
\end{equation}

This problem was solved for a non-zero initial temperature in \cite{AgeevAref:Waking,AgeevAref:Memory} and the answer is given by
\begin{equation}\label{S-AA}
	S(\ell,t) = \log  \left( \frac{z_h} {\ell \,\fS_\kappa (s,\rho)}\,\sinh\frac{t}{z_h}\right), \hspace{1cm}
	0 \leq t \leq \ell,
\end{equation}
where $\fS_\kappa$ is
\begin{equation}\label{fS}
	\fS_\kappa (s,\rho) = \frac {c \rho + \Delta} {\Delta}\cdot\sqrt {\frac {\Delta^2 - 
		c^2 \rho^2} {\rho\left (c^2 \rho + 2 c\Delta + \rho \right) - \kappa^2}},
\end{equation}
and the following notations are used
\begin{equation}\label{Smin}
	\kappa = \frac{z_h}{z_H}<1,\,\,\,\,
	c = \sqrt{1-s^2},\,\,\,\,
	\Delta = \sqrt{\rho^2-\kappa^2},\,\,\,\,
	\gamma = 1-\kappa^2.
\end{equation}

The function $\fS_\kappa$ depends on the new variables $s,\rho$ which are related to the variables $\ell,t$. The new variables $s,\rho$ describe the geodesic in the bulk space and are expressed in terms of the position of the geodesic top $z_*$ and the point $z_c$ where the geodesic crosses the lightlike shell by the formulas
\begin{equation}\label{s-rho}
	s = \frac{z_c}{z_*}, \hspace{1cm} \rho=\frac{z_h}{z_c},
\end{equation}
with the restrictions $z_*<z_H$ and $z_c<z_H$ satisfied.

Variables $\ell,t$ are related to $s,\rho$ by the following expressions
\begin{align}
	\label{ell}	\ell &= \frac{z_h}{2}  \log \frac{c^2 \gamma ^4-4 \Delta \left(c s \left(\kappa ^2-2 \rho ^2+1\right) + \Delta +\Delta \left(\rho ^2-2\right) s^2\right)} {c^2 \gamma^4-4 \Delta ^2 (\rho  s-1)^2}+\frac{z_h}{2 \kappa } \log\frac{(c \kappa +\Delta s)^2}{\rho ^2 s^2-\kappa^2}, \\
	\label{ttt} t &= z_h\, {\mbox {arccoth}}\left(\frac{-c \kappa ^2+2 c \rho
	^2+c+2 \Delta  \rho }{2 c \rho +2 \Delta }\right).
\end{align}

We are interested in calculating $S=S(\ell,t)$. Hence, we have to find $\rho=\rho(\ell,t)$ and $s=s(\ell,t)$ first. From equation \eqref{ttt} it is possible to find $\rho=\rho(s,t)$. Inserting this expression for $\rho$ into \eqref{ell}, we obtain $\ell=\ell((s,\rho(s,t))=\ell(s,t)$, from which we get numerically $s=s(\ell,t)$ and $\rho=\rho(s(\ell,t),t)=\rho(\ell,t)$. And finally, we can get $S=S(s(\ell,t),\rho(\ell,t))=S(\ell,t)$.

\vspace{1cm}
\section{Entropy quantities of a multipartite system}

\subsection{Mutual information}

Recall that for an arbitrary bipartite quantum state $\rho_{AB}$ the \textit{quantum mutual information} is defined as (see, for example \cite{OhyaVol,NielsenChuang})
\begin{equation}\label{QMI}
	I(A:B) = S(\rho_A) + S(\rho_B) - S(\rho_{AB}),
\end{equation}
where $S(\rho_A) = -\tr(\rho_A \log\rho_A)$ is the entanglement entropy, 
$\rho_A=\tr_B(\rho_{AB})$ is the reduced density matrix.
It measures the total amount of correlation (both classical and quantum) between $A$ and $B$, as measured by the minimal rate of randomness that are required to completely erase all the correlations in $\rho_{AB}$ \cite{GroisPopWint}. 
In Fig.\:\ref{fig:Venn-diag}, the mutual information of $A$ and $B$ corresponds to $b+d$.

If all terms in the RHS of \eqref{QMI} are calculated holographically in accordance with previous section then we get the \textit{holographic mutual information} for a two-party system
\begin{equation}\label{HMI}
	I(A:B) = S(A) + S(B) - S(AB),
\end{equation}
where $S(A)$ is the Ryu--Takayanagi (holographic) entropy of segment $A$, $AB$ denotes $A \cup B$, $S(AB)$ is the holographic entanglement entropy for the union of two subsystems.

It was proved (e.g. \cite{Ruelle,ArakiLieb}) that the mutual information is always non-negative: 
\begin{equation}\label{}
	I(A:B) \geqslant 0.
\end{equation}
This property is called \textit{subadditivity} of entropy.

\subsection{Tripartite information}
\begin{figure}
	\centering
	\includegraphics[scale=0.5]{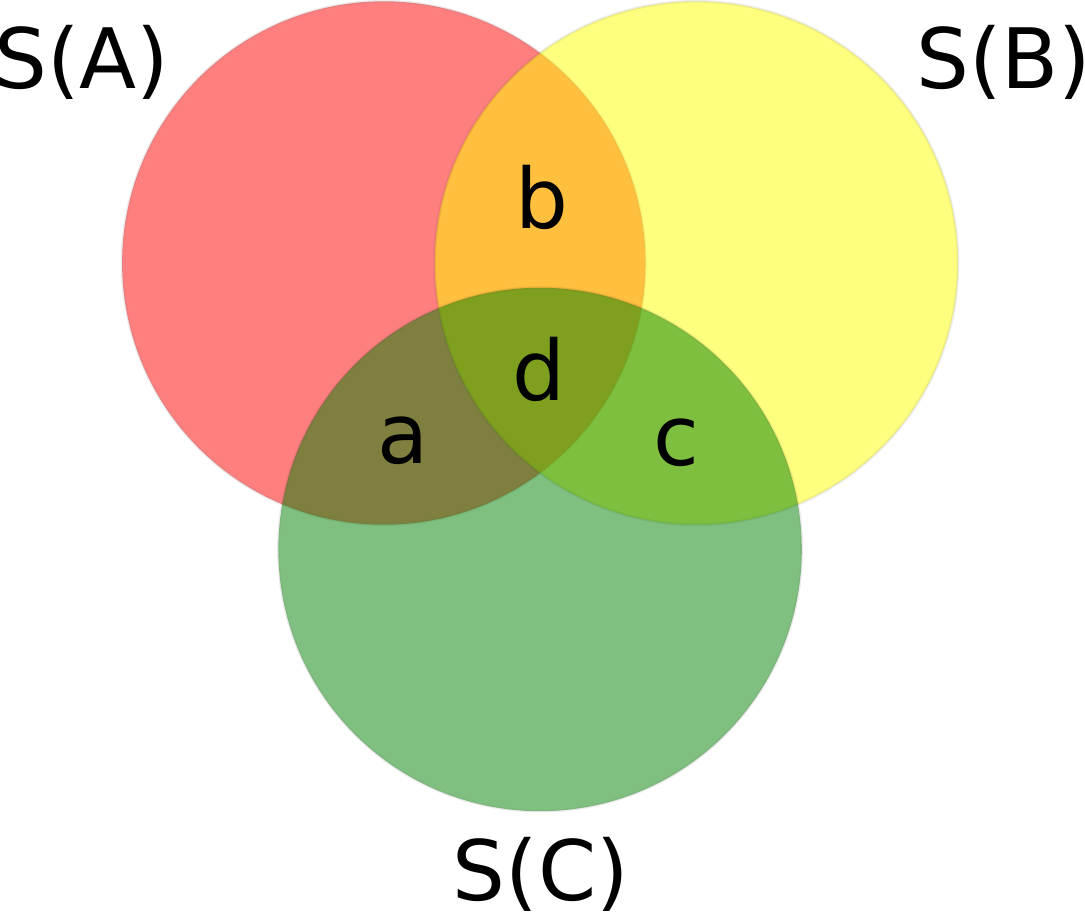}
	\caption{Venn diagram for a three-party system. $I(A:B)=b+d$, $TI(A:B:C)=d$, $TC(A:B:C)=a+b+c+2d$, $SM(A:B:C)=a+b+c+d$.}
	\label{fig:Venn-diag}
\end{figure}

The mutual information of a three-party system (\textit{tripartite information}, \textit{interaction information}) is usually defined as
\begin{align}
	TI(A:B:C) = I(A:B) + I(A:C) - I(A:BC) = \\
	= S(A)+S(B)+S(C) - \left[S(AB)+S(BC)+S(AC)\right] + S(ABC),
\end{align}
where $ABC$ denotes $A \cup B \cup C$.
In Fig.\:\ref{fig:Venn-diag}, the tripartite information of $A, B, C$ corresponds to $d$.

As it was shown by Casini and Huerta \cite{CasiniHuerta}, in a general quantum system the tripartite information can be either positive, negative, or zero depending on the choice of $A, B, C$.

Hayden et al. proved \cite{HaydHeadMal:Monogam} that for any regions $A,B,C$ the holographic tripartite information is non-positive:
\begin{equation}\label{}
	TI(A:B:C) \leqslant 0.
\end{equation}

This inequality can be written in the form
\begin{equation}\label{monogamy}
	I(A:B) + I(A:C) \leqslant I(A:BC).
\end{equation}
Inequalities that are structurally of the same form as \eqref{monogamy} appear frequently in quantum information theory and quantum cryptography and are associated with the \textit{monogamy} relations. And the holographic tripartite information is said to be \textit{monogamous} \cite{HaydHeadMal:Monogam}.

The question arises how is this negative value of $TI$ useful for quantum information tasks? There is an opinion \cite{Kumar} that the fact that tripartite information, being a measure of correlation, can assume negative value is challenging.

\subsection{Total correlation}

The \textit{total correlation} is a straightforward generalization of the mutual information to a multipartite case and is defined by
\begin{equation}\label{TC}
	TC(A_1:A_2:\ldots:A_m) = \sum_{i=1}^m S(A_i) - S(A_1A_2\ldots A_m).
\end{equation}
This quantity was used, for example, in \cite{Horodecki,GroisPopWint,Yang:Squashed,Shir:Measures,Watanabe} and also in \cite{CerfMassarSchn:secr_monot} where it is called the \textit{second secrecy monotone}.
In Fig.\:\ref{fig:Venn-diag}, the total correlation of $A, B, C$ corresponds to $a+b+c+2d$.

The total correlation can be represented as a sum of bipartite mutual information \cite{CerfMassarSchn:secr_monot}:
\begin{equation}\label{key}
	TC(A_1:A_2:\ldots:A_m) = I(A_1:A_2) + \sum_{i=2}^{m-1} I(A_1\ldots A_i:A_{i+1}).
\end{equation}
Since the mutual information is non-negative, the total correlation is also non-negative:
\begin{equation}
	TC(A_1:A_2:\ldots:A_m) \geqslant 0.
\end{equation}

\subsection{Secrecy monotone}

In 1978 Han suggested \cite{Han} the entropy measure called \textit{dual total correlation} for classical random variables.
Studying multipartite quantum cryptography, Cerf et al. \cite{CerfMassarSchn:secr_monot} introduced a quantity called \textit{secrecy monotone} which is, in fact, quantum generalization of the dual total correlation. 
Later, the same quantity was independently proposed and analyzed by Kumar \cite{Kumar}. 
This quantity was also used in \cite{Yang:Squashed,Shir:Measures,SazimAgrawal}.
In literature it is sometimes called \textit{binding information}, or \textit{operational quantum mutual information}.

The secrecy monotone of an $m$-party system is defined by
\begin{equation}\label{}
	SM(A_1:A_2:\ldots:A_m) = \sum_{i=1}^m S(A_1 \ldots \check{A_i} \ldots A_m) - (m-1)S(A_1A_2\ldots A_m),
\end{equation}
where $\check{A_i}$ means that $A_i$ is omitted.

In particular, the secrecy monotone for a three-party system is
\begin{equation}\label{}
	SM(A:B:C) = S(AB) + S(BC) + S(AC) - 2S(ABC).
\end{equation}
In Fig.\:\ref{fig:Venn-diag}, the secrecy monotone of $A, B, C$ corresponds to $a+b+c+d$.

It was shown in \cite{CerfMassarSchn:secr_monot,Kumar} that the secrecy monotone is non-negative:
\begin{equation}\label{}
	SM(A_1:A_2:\ldots:A_m) \geqslant 0.
\end{equation}

For two parties, the secrecy monotone is, obviously, equal to the mutual information 
\begin{equation}\label{}
	SM(A:B) = S(AB) + S(B) + S(A) - 2S(AB) = I(A:B).
\end{equation}

\vspace{1cm}
\section{Evolution of the entanglement entropy\\ quantities}

\begin{figure}[H]
	\centering
	\includegraphics[scale=0.45]{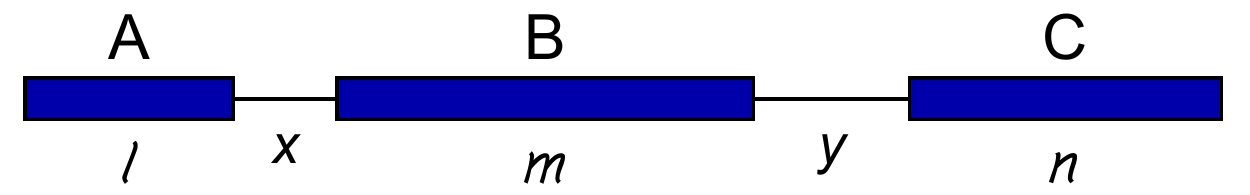}
	\caption{A three-segment system of our interest.}
	\label{fig:3segm}
\end{figure}

In this section, we investigate time dependence of the above quantities for a system consisting of three segments. Let us denote lengths of the segments as $l, m, n$ and distances as $x, y$ (see Fig.\:\ref{fig:3segm}).

\subsection{Tripartite information}

\begin{figure}[h]
	\centering
	\includegraphics[scale=0.5]{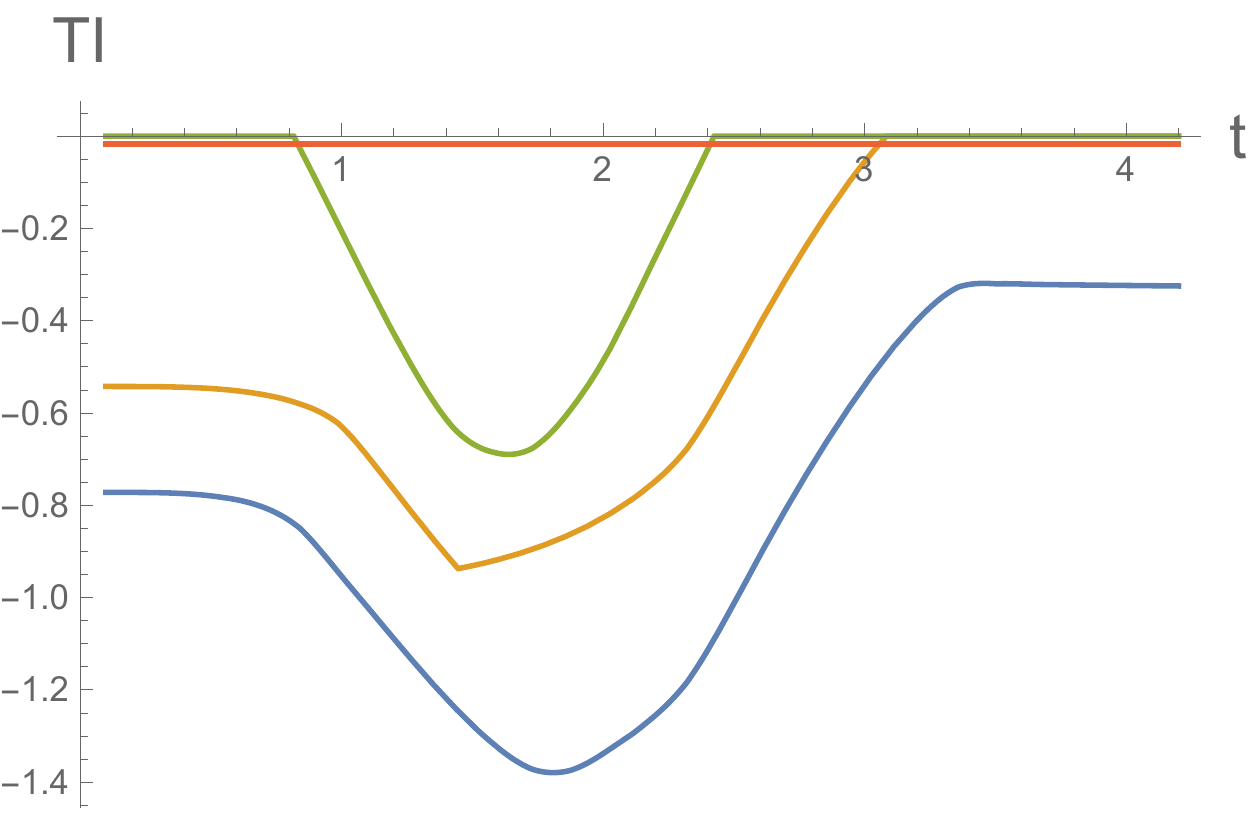}
	\caption{Four types of time dependence of the holographic tripartite information.}
	\label{fig:ti-4forms}
\end{figure}

We have found the following four types of time dependence of the holographic tripartite information  (Fig.~\ref{fig:ti-4forms}) 

1) Wake-up and scrambling times are absent, and the tripartite information is always negative;

2) Wake-up time is absent, but scrambling time is present;

3) Wake-up and scrambling times are present, and a plot of the tripartite information has an upturned bell shape;

4) The tripartite information is identically equal to zero.

The specific plot type of the holographic tripartite information depends on all geometrical parameters of the system (i.e. lengths of the segments and distances between them) and on the temperature. In particular, by increasing $z_H$, increasing the lengths of the segments, decreasing the distances between them, the absolute value of the holographic tripartite information increases.

The same types of time dependence of the holographic tripartite information were previously obtained in \cite{AlishMozTanh}.

\subsection{Total correlation and secrecy monotone}

\begin{figure}[H]
	\centering
	\includegraphics[scale=0.5]{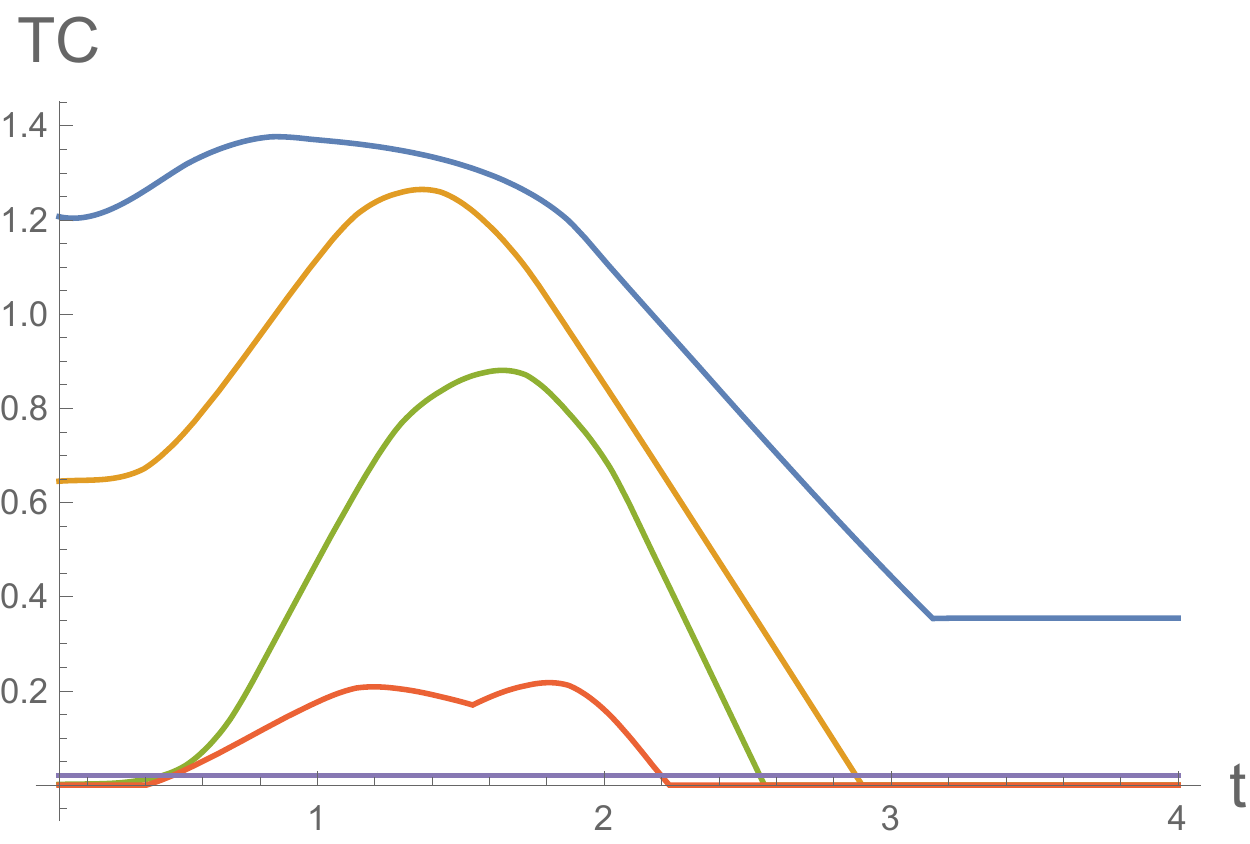}A \hspace{5mm}
	\includegraphics[scale=0.5]{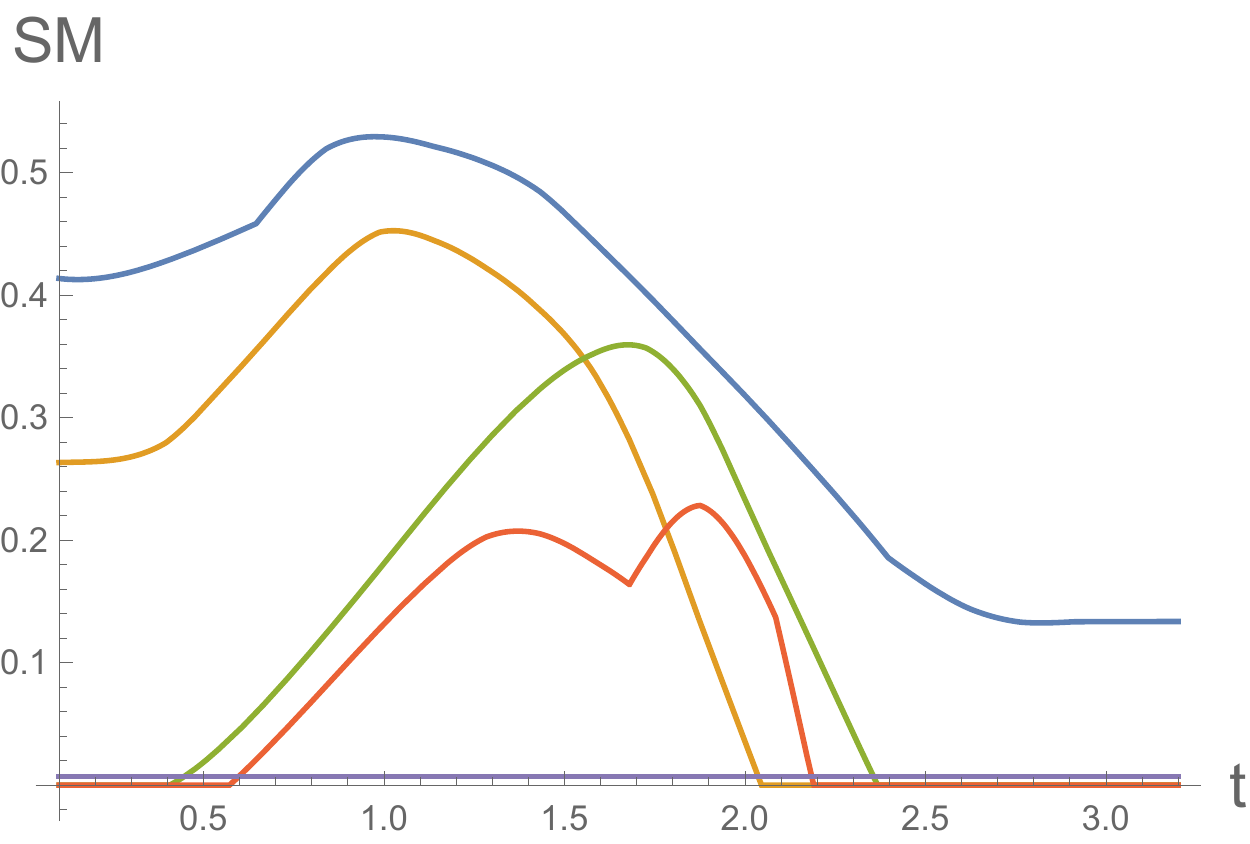}B
	\caption{Five types of time dependence of the holographic total correlation (A) and the holographic secrecy monotone (B).}
	\label{fig:tc_sm-5forms}
\end{figure}

We have found the following five types of time dependence of the holographic total correlation (Fig.~\ref{fig:tc_sm-5forms}.A) and the holographic secrecy monotone (Fig.~\ref{fig:tc_sm-5forms}.B)

1) Wake-up and scrambling times are absent, and both quantities are always positive;

2) Wake-up time is absent, but scrambling time is present;

3) Wake-up and scrambling times are present, and plots of both quantities have a bell shape;

4) Wake-up and scrambling times are present, and plots of both quantities have a two-hump shape;

5) Both quantities are identically equal to zero.

The specific plot type of the holographic total correlation and the holographic secrecy monotone depends on all geometrical parameters of the system (i.e. lengths of the segments and distances between them) and on the temperature. In particular, by increasing $z_H$, increasing the lengths of the segments, decreasing the distances between them, the holographic total correlation and the holographic secrecy monotone increase.

Note that these five types of time dependence coincide exactly with the general types of time dependence of the holographic mutual information for a composite quantum system in \cite{ArefInozVol:Hologr}. 

It is interesting that the similar time dependence of the mutual information was found in the field of quantum biology. Bradler et al. \cite{Bradler} simulated quantum effects in photosynthetic light-harvesting complexes using Gorini–Kossakowski–Sudarshan–Lindblad master equation (i.e. without holography) and found a bell shape and a two-hump shape of the mutual information behavior.

\begin{figure}[!ht]
	\centering
	\includegraphics[scale=0.5]{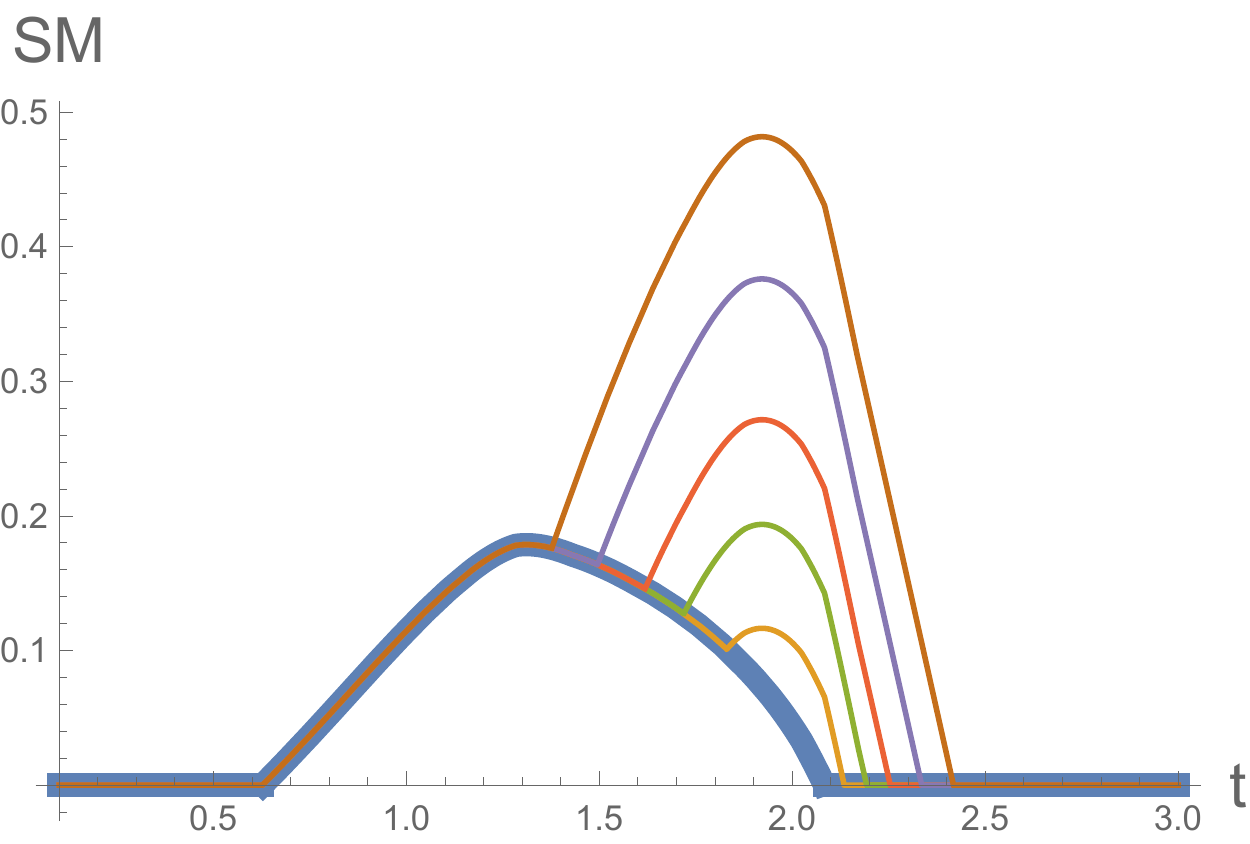}A \hspace{7mm}
	\includegraphics[scale=0.5]{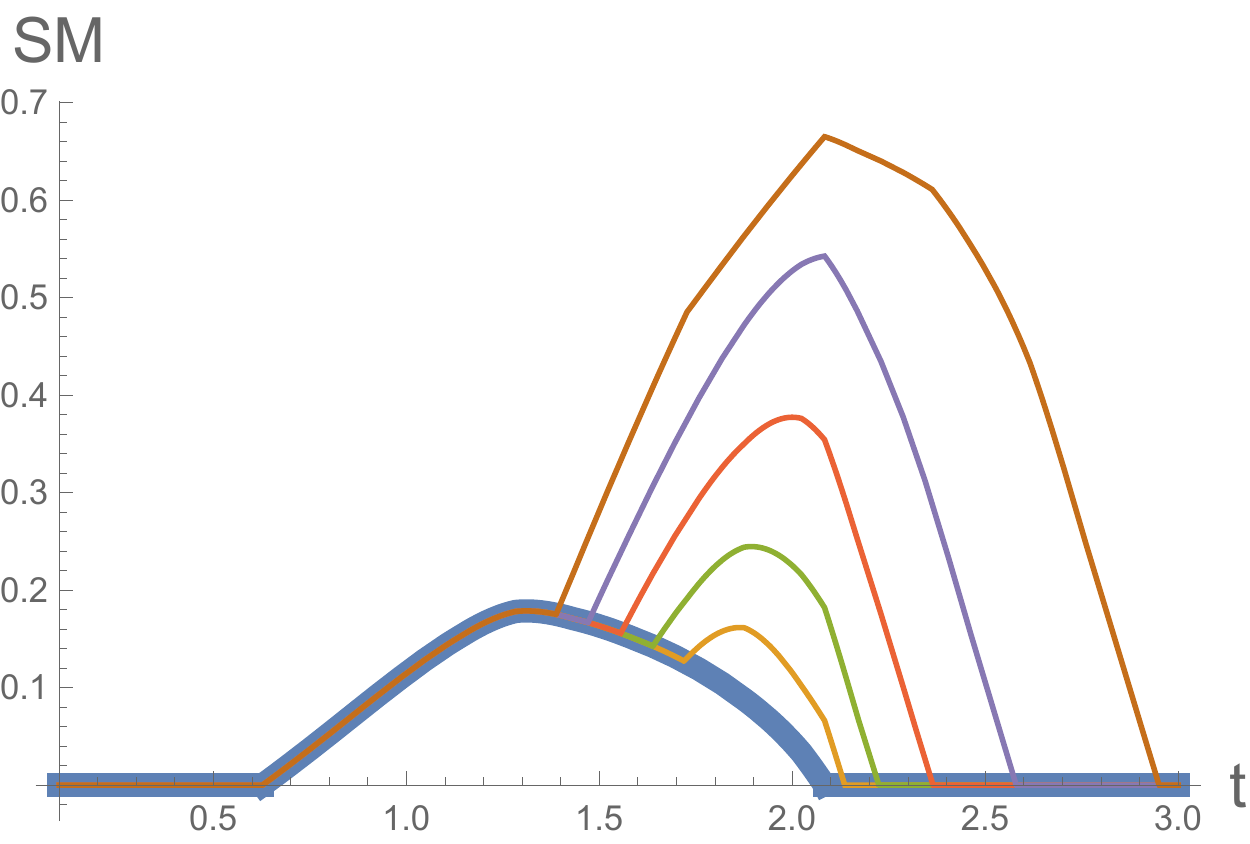}B
	\caption{Origination of a two-hump shape of the holographic secrecy monotone for a three-segment system.}
	\label{fig:SM-bell}
\end{figure}

Fig.\:\ref{fig:SM-bell} shows the origination of a two-hump shape of the holographic secrecy monotone for a three-segment system (Fig.\:\ref{fig:3segm}).
Let $A$ and $B$ be two such segments that the corresponding plot of the holographic mutual information has a bell shape (thick blue curve). Fig.\:\ref{fig:SM-bell}.A shows the origination of the second hump as segment $C$ of fixed length approaches segments $A$ and $B$ from afar right (as $y$ decreases). Fig.\:\ref{fig:SM-bell}.B shows the origination of the second hump as the length of segment $C$ increases with the distances fixed (as $n$ increases).

\vspace{1cm}
\section{Conclusions}

Using the holographic approach, we studied the evolution of the tripartite information, total correlation and secrecy monotone for a three-party system. We have shown how to control these quantities by changing geometrical parameters of the system and its temperature.

We have found the following four types of time dependence of the holographic tripartite information \vspace{2mm}

1) wake-up and scrambling times are absent, and the tripartite information is always negative;

2) wake-up time is absent, but scrambling time is present;

3) wake-up and scrambling times are present, and a plot of the tripartite information has an upturned bell shape;

4) the tripartite information is identically equal to zero; \vspace{2mm} \\
and the following five types of time dependence of the holographic total correlation and the holographic secrecy monotone
\vspace{2mm}

1) wake-up and scrambling times are absent, and both quantities are always positive;

2) wake-up time is absent, but scrambling time is present;

3) wake-up and scrambling times are present, and plots of both quantities have a bell shape;

4) wake-up and scrambling times are present, and plots of both quantities have a two-hump shape;

5) both quantities are identically equal to zero.

\vspace{2mm}
We have shown that by changing the system parameters one can obtain a bell shape and a two-hump shape of the holographic total correlation and the holographic secrecy monotone. Bell shape and, especially, two-hump shape realize with quite small ranges of system parameters. We have also explained the second hump origination on the holographic secrecy monotone plot.

\vspace{1cm}
\subsection*{Acknowledgments}

This work was supported by the Russian Science Foundation (project \textnumero14-50-00005).

We thank Mark M. Wilde for bringing the paper \cite{Han} to our attention.

\end{document}